\documentclass[aps,pre,superscriptaddress,unsortedaddress]{revtex4}
\usepackage{amsmath,amsfonts,amssymb}
\usepackage[english]{babel} 
\usepackage[latin1]{inputenc} 
\usepackage[T1]{fontenc}
\usepackage{color}
\usepackage{float}
\usepackage{verbatim}
\usepackage[pdftex]{graphicx}
\usepackage{bm}
\usepackage{mathtools}
\usepackage{tikz}

\definecolor{linkcolor}{rgb}{0,0,0.6} %hyperlink
\usepackage[pdftex,colorlinks=true, pdfstartview=FitV, linkcolor= linkcolor, citecolor= linkcolor, urlcolor= linkcolor, hyperindex=true,hyperfigures=true]{hyperref} %hyperlink

\begin{document}

\title{Engineered Swift Equilibration for Brownian objects: from underdamped to
overdamped dynamics}

\author{Marie Chupeau}
\affiliation{LPTMS, CNRS, Univ. Paris-Sud, Universit\'e Paris-Saclay, UMR 8626, 91405 Orsay,
  France}
  
\author{Sergio Ciliberto}
\affiliation{Université de Lyon, CNRS, Laboratoire de Physique de l'\'Ecole Normale Sup\'erieure, UMR5672, 46 All\'ee d'Italie, 69364 Lyon, France.}

\author{David Gu\'ery-Odelin}
\affiliation{Laboratoire de Collisions Agr\'egats R\'eactivit\'e, CNRS, UMR 5589, IRSAMC, France }

\author{Emmanuel Trizac}
\affiliation{LPTMS, CNRS, Univ. Paris-Sud, Universit\'e Paris-Saclay, UMR 8626, 91405 Orsay,
  France}

\begin{abstract}
We propose a general framework to study transformations that drive an underdamped Brownian particle in contact with a thermal bath from an equilibrium state to a new one in an arbitrarily short time.
To this end, we  
make use of a time and space-dependent potential, that plays a dual role: confine the particle, and manipulate the system. In the special case of an isothermal compression 
or decompression of a harmonically trapped particle, we derive explicit 
protocols that perform this quick transformation, following an inverse engineering method. We focus on the properties of these protocols, which crucially depend on two key dimensionless numbers 
that characterize the relative values of the three timescales of the problem, associated with friction, oscillations in the confinement and duration of the protocol. 
In particular, we show that our protocols encompass the known overdamped version of this problem and extend it to any friction for decompression and to a large range of frictions for 
compression. 
\end{abstract}

\maketitle

\section{Introduction}

Shortcuts To Adiabaticity (STA) 
emerged in quantum mechanics as fast protocols for state-to-state transformations that would otherwise 
require the slow and therefore time-consuming modification of a control 
parameter of the system to reach the desired final state following a quasi-adiabatic trajectory~\cite{Torrontegui2013_117}. Many strategies have been proposed to set up non-adiabatic routes to reach the same final state through the use of dynamical invariants \cite{Chen2010_063002}, 
counter adiabatic driving \cite{Demirplak2003,Berry2009,An:2016}, reverse engineering methods \cite{Muga2009,PhysRevA.83.013415,Zhang2017}, fast-forward techniques \cite{Masuda2008,Masuda2009}, 
Lie algebraic approaches \cite{Martinez-Garaot2014_053408,Torrontegui2014}, and optimal control \cite{Stefanatos2010,Chen2011_043415,GlaserPRA,GlaserEPJD} to name but a few. Slow processes (adiabatic in quantum mechanics jargon)
and thus STA are quite common to prepare the state of the system in a wide variety of domains including atomic and molecular physics \cite{Bason2012,Du2016}, quantum transport 
\cite{Couvert2008, Bowler2012,Walther2012}, solid state \cite{Zhou2017_330}, many-body physics \cite{delCampo2013,Rohringer2015,Deng:2018}, classical mechanics \cite{Deffner2014} and 
statistical physics \cite{Guery:2014,Martinez:2016b,LeCunuder:2016}. STA also have applications in the design of optimal devices, as recently proposed in optics \cite{Ho2015} and 
in internal state manipulation for interferometry \cite{Impens2017}.
STA  therefore enjoy a large domain of applications, and the number of experiments demonstrating their efficiency is soaring.   

Recently, these techniques have given birth to new protocols in statistical physics. Thermodynamic transformations that connect two different equilibrium states are not in most cases quasi-static
 and thus necessarily visit out-of-equilibrium states. Operating such transformations in a finite and short amount 
of time, potentially much shorter than the relaxation time of the system, is crucial for many applications, in particular in micro and nano devices or engines
\cite{Kaka:2005,Blickle:2012,Dechant:2015,Rossnagel:2016,Martinez:2016c,Dechant:2017}, triggering a number of works considering how STA could boost engines, among which~\cite{Deng:2013,Campo:2014,Tu:2014,Beau:2016}. As in quantum 
physics, performing this kind of quick transformations requires to devise an appropriate driving of the intermediate out-of-equilibrium dynamics. A recent example has been 
provided with protocols to compress or decompress an isolated 3D harmonically trapped cloud of atoms in an arbitrarily short amount of time \cite{Guery:2014}. More importantly, 
such an approach has been generalized to systems in contact with a thermostat which are ubiquitous in thermodynamics. The so-called Engineered Swift Equilibration (ESE) 
protocols have been introduced to study the isothermal compression of a colloidal particle~\cite{Martinez:2016b}. The idea 
was then adapted to encompass the shift of the cantilever of an Atomic Force Microscope~\cite{LeCunuder:2016}, and 
has been generalized in~\cite{Li:2017} 
to underdamped processes using a non-conservative driving force. 
Here, our interest also goes to the underdamped dynamics of a Brownian particle,
under the proviso that the driving force, used to manipulate the system but also to confine
the particle, is conservative. 

In section \ref{sec2}, we address the general case of ESE protocols in the underdamped regime, for a non-isothermal transformation and generic potentials. 
Contrarily to the works~\cite{Tu:2014,Li:2017}, we resort to conservative drivings, through potentials that only depend on space and not on velocity. This general framework 
leads to lengthy equations, that are significantly simplified in the case of transport-free harmonic potentials. We next restrict to this class of transformations 
in section~\ref{sec3} and show how to obtain a fully explicit isothermal protocols, choosing the shape of one characteristic quantity of the particle density function and 
deducing from it the appropriate evolution of the control parameter. Finally, in section~\ref{sec4} we exhibit and analyze thoroughly the ``phase diagram'' of such protocols. 
We proceed to show that it largely depends on whether the transformation is a compression or a decompression, and work out the various properties of these protocols, such 
as existence, cross-over to the overdamped regime, 
position-velocity decoupling or also transient negativity of the stiffness. We supplement this by a discussion on the shape of the temporal evolution of the stiffness, through the comparison between the relevant timescales of the problem.
We also analyze the robustness of the phase diagrams with respect to the shape of the protocol, and comment on the change occurring in the protocol when its duration is decreased. We conclude in section \ref{sec:concl}.

%%%%%%%%%%%%%%%%%%%%%%%%%%%%%%%%%%%%%%%%%%%%%%%
\section{General formalism}
\label{sec2}

The ESE protocol brought to the fore in \cite{Martinez:2016b} addressed the case of an overdamped confined Brownian object.
While the overdamped limit is suited for colloids in a solvent like water, it is desirable to 
study the generalization of the idea to underdamped situations, when inertial effects no longer are negligible, such as for an Atomic Force Microscope tip where friction is on purpose reduced as much as possible~\cite{Giessibl:2003}, or for the study of a levitated nanoparticle in air where friction can be tuned through gas pressure~\cite{Rondin:2017}. 
Generically, when viscous friction is not high compared to the other characteristic frequencies of the problem,
one should include the velocity degrees of freedom in the description in addition to positional ones; the overdamped approximation, on the other hand, assumes that the former are equilibrated at all times. 
To extend the ESE method proposed in~\cite{Martinez:2016b} to the underdamped description of an object immersed in a thermal bath trapped in a confining potential,
we introduce the probability density function $K(x,v,t)$ of the position $x$ and velocity $v$ of the particle. It obeys the Kramers equation

\begin{equation}\label{kramers}
\partial_t K + v \partial_x K -\frac{\partial_x U}{m} \partial_v K = \gamma \partial_v (v K) + \gamma \frac{k_B T}{m} \partial^2_v K
\end{equation}
where $U(x,t)$ is the confining potential, $m\gamma$ the damping coefficient in the fluid with $m$ the mass of the particle, $k_B$ the Boltzmann constant and $T$ the 
temperature of the bath. At thermal equilibrium, this probability density function is simply given by the Boltzmann law
\begin{equation}
K_{\rm eq}(x,v)=K_0 \exp\left( - \frac{U(x)}{k_B T}-\frac{m v^2}{2k_B T} \right).
\end{equation}
To connect an initial equilibrium state characterized by the potential $U_i(x)$ and the temperature $T_i$ to a final equilibrium state ($U_f(x)$, $T_f$), we assume 
that the probability density function keeps a Gaussian form in $v$ during the transformation
\begin{equation}\label{ansatzgen}
K(x,v,t)=\exp(-{\cal A}(x,t)-{\cal B}(x,t) v^2-{\cal D}(x,t) v).
\end{equation}
This ansatz, inspired by previous works on Boltzmann equation where it results from the use of Boltzmann $H$-theorem~\cite{Guery:2014},
remains operational here. 
In Eq.~\eqref{ansatzgen}, $1/{\cal B}$ plays the role
of a kinetic temperature, that should at equilibrium coincide with that of the bath ($T$), but is otherwise distinct.
It is worth emphasizing here that the bath temperature can be time-dependent. In the colloidal realm for example, this
is achieved by an appropriate random shaking of the confinement potential \cite{ber14,Chupeau:2018a}, which creates an effective
temperature for the Brownian object while the {\em true}
bath temperature remains constant.

In the spirit of ESE techniques, we do not impose the control function/parameter $U(x,t)$ and $T(t)$ beforehand to study the response of the system through the functions 
${\cal A}$, ${\cal B}$ and ${\cal D}$. On the contrary, we adopt a reverse engineering point of view, namely we choose a desired dynamics for these functions and 
deduce from it the temporal evolution of the control parameters that needs to be enforced to perform the chosen dynamics. Functions ${\cal A}$, ${\cal B}$, ${\cal D}$ 
and control parameters $U$ and $T$ are linked via a set of equations that we obtain by plugging the ansatz form~\eqref{ansatz} into the Kramers equation~\eqref{kramers} 
and sorting the $v^3$, $v^2$, $v$ and constant terms
\begin{subequations}\label{eqESEgen}
\begin{align}
& \partial_x {\cal B}=0 \label{Bgen}\\
&-\dot{\cal B}-\partial_x {\cal D}=2\gamma \left(\frac{2 {\cal B}^2 k_B T}{m} -{\cal B}   \right) \label{Bgen2}\\
&-\dot{\cal D}-\partial_x {\cal A} +2\frac{\partial_x U}{m}{\cal B}=\gamma \left(-{\cal D} +4 \frac{k_B T}{m} {\cal B} {\cal D} \right) \label{Dgen}\\
&- \dot{\cal A}+\frac{\partial_x U}{m}{\cal D} = \gamma \left(1+\frac{{\cal D}^2k_B T}{m} -\frac{2 {\cal B} k_B T}{m} \right) \label{Agen}
\end{align}
\end{subequations}
where the dot stands for time derivative. This set of equations, general within the ansatz~\eqref{ansatzgen}, must be obeyed to connect the two imposed equilibrium states. Of course, they have to be adapted to specific protocols with desired constraints, such as duration of the protocol,  amplitude of the transition and number and nature of control parameters. The knowledge of the existence of at least one of these specific protocols is in general not a simple task, as we will see later. However, let us focus on these equations and try to extract as much information as we can. They describe
Gaussian compressions and decompressions with transport when $U(x,t)$ 
remains quadratic, but the $x$ part of the particle density function can also stray from the Gaussian shape when arbitrary potentials are used. For clarity and simplicity 
purposes, we will restrict ourselves to harmonic potentials of angular frequency $\omega(t)$ (stiffness $\kappa=m\omega^2$) and center position $x_0(t)$
\begin{equation}
U(x,t) \, = \, \frac{m \omega^2(t)}{2}\, \left[ x-x_0(t) \right]^2,
\label{eq:harmonic}
\end{equation}
so as to carry out a compression or a decompression from the initial state characterized by $(\omega_i,T_i)$ to the final state $(\omega_f,T_f)$. With this harmonic potential, our Kramers equation is linear in $x$ and $v$, ensuring that the particle density function keeps a Gaussian shape at all times. A first analysis of equations~\eqref{Bgen} and~\eqref{Bgen2} shows that 
the kinetic temperature $1/{\cal B}$ is  
time-dependent only and does not depend on space, and that ${\cal D}$ is linear in $x$
\begin{equation}\label{D2}
{\cal D}(x,t)= \left[ -\dot{{\cal B}} + 2\gamma \left( {\cal B}-2{\cal B}^2\frac{k_B T}{m} \right) \right]x+{\cal D}_0(t).
\end{equation}
Integrating equation~\eqref{Dgen} with respect to $x$ then yields a quadratic form in $x$ for ${\cal A}(x,t)$
\begin{align}\label{A2}
{\cal A}(x,t)=& \left[ \ddot{\cal B} -3\gamma \dot{\cal B}+12 {\cal B} \dot{\cal B} \frac{\gamma k_B T}{m} + 4 \gamma {\cal B}^2 \frac{k_B \dot{T}}{m} +2\omega^2{\cal B}+2\gamma^2 {\cal B} -12\gamma^2 {\cal B}^2\frac{k_B T}{m} + 16 \left( \frac{\gamma k_B T}{m} \right)^2 {\cal B}^3 \right] \frac{x^2}{2} \nonumber \\
&- 2\omega^2 {\cal B}x x_0- \dot{\cal D}_0 x - \gamma \left( 4 {\cal B} \frac{k_B T}{m} -1 \right) {\cal D}_0 x +\omega^2{\cal B} x_0^2 +{\cal A}_0(t)
\end{align}
where the function ${\cal A}_0(t)$ is related to the normalisation of the distribution. Its temporal evolution can be left aside from our study, as our Fokker-Planck equation 
conserves probability during the transformation. If we plug expressions~\eqref{D2} and~\eqref{A2} into equation~\eqref{Agen}, and sort out the monomials in $x$, we obtain 
the equations controlling the time evolution of ${\cal B}(t)$
\begin{align}
&\overset{\ldots}{\cal B}-3\gamma \ddot{\cal B} + 12\frac{\gamma k_B T}{m} \left( \dot{\cal B}^2 + {\cal B} \ddot{\cal B} \right) +20 {\cal B} \dot{\cal B} \frac{\gamma k_B \dot{T}}{m}+ 4 \gamma {\cal B}^2 \frac{k_B \ddot{T}}{m} + 4\omega \left(\dot{\omega} {\cal B}+\omega \dot{\cal B}-\gamma \omega{\cal B}\right)  \nonumber \\
&+ 2 \gamma^2 \dot{\cal B}- 32 \gamma^2 {\cal B} \dot{\cal B} \frac{k_B T}{m} - 12 \gamma^2 {\cal B}^2 \frac{k_B \dot{T}}{m}+32 \gamma^2 k_B^2 \frac{T \dot{T}}{m^2} {\cal B}^3 + 64 \left( \frac{\gamma k_B T}{m} \right)^2 {\cal B}^2 \dot{\cal B}\nonumber \\
&+ 8 \omega^2 {\cal B}^2 \frac{\gamma k_B T}{m} + \frac{2 \gamma k_B T}{m} \left[ \dot{\cal B}^2+4\gamma^2 {\cal B}^2 + 16 {\cal B}^4 \gamma^2 \left( \frac{k_B T}{m} \right)^2-16 \gamma^2 {\cal B}^3 \frac{k_B T}{m} \right]=0
\end{align}
as well as the time evolution of ${\cal D}_0(t)$
\begin{align}\label{D0}
& \ddot{\cal D}_0+\gamma \left( 4\frac{{\cal B} k_B T}{m}-1\right) \dot{\cal D}_0+\left[\omega^2 +2 \frac{\gamma k_B}{m} \left( 3 \dot{\cal B}T+2 {\cal B} \dot{T} - 2 \gamma {\cal B} T+4 {\cal B}^2 \gamma \frac{k_B T^2}{m} \right) \right] {\cal D}_0 \nonumber \\
  &= x_0 \left( 4 {\cal B} \omega \dot{\omega} +3 \omega^2 \dot{\cal B} +2 \gamma {\cal B} \omega^2 -4 \omega^2 {\cal B}^2 \frac{\gamma k_B T}{m} \right) +2\omega^2 {\cal B} \dot{x}_0.
\end{align}
Both equations are informative. The first one is not very tractable in itself, but highlights that the inverse kinetic temperature ${\cal B}$ is completely determined by the bath temperature $T$ and 
the trap stiffness $\omega$, and thus independent of the transport part $x_0(t)$ of the transformation. On the other hand, the second equation tells us that ${\cal D}_0(t)$ 
is fully induced by transport. With no transport, $x_0(t)$ {vanishes during the whole transformation, and so does ${\cal D}_0(t)$ since initially ${\cal D}_0(0)=0$. 
As a result,} ${\cal D}(x,t)$ is simply proportional to $x$, 
referring to equation~\eqref{D2}. Finally, coming back to equation~\eqref{A2}, in a transport-free case, ${\cal A}(x,t)$ is itself simply proportional to $x^2$.

%%%%%%%%%%%%%%%%%%%%%%%%%%%%%%%%%%%%%%%%%%%%%%%%%%%%%%%%%%%%%%%%%%%%%%%%%%%%%%%%%%%%%%%%%%
\section{Harmonic transport-free protocol}
\label{sec3}

\subsection{Simplified formalism}

We now restrict to transport-less transformations, carried out by a harmonic potential. A variant of this problem was numerically solved in~\cite{Gomez:2008}, and the corresponding protocol displayed discontinuities at initial and final times. Here, we will provide an exact explicit solution while imposing smooth boundary conditions, in order to create a protocol well-adapted to experiments. Our Kramers ansatz can be written in the lighter form
\begin{equation}\label{ansatz}
K(x,v,t)=N(t)\exp(-\alpha(t)x^2-\beta(t) v^2-\delta(t)x v),
\end{equation}
with the correspondence $ {\cal A}(x,t)=\alpha(t)x^2-\ln N$, where $N(t)$ is a normalization factor, $ {\cal B}(x,t)=\beta(t)$ and ${\cal D}(x,t)=\delta(t) x$, the function 
$\delta(t)$ being the amplitude of $x-v$ correlations. The set of equations~\eqref{eqESEgen} then comes down to
\begin{subequations}\label{eqESE}
\begin{align}
&- \dot{\alpha}+\omega^2 \delta = \frac{\gamma k_B T}{m} \delta^2 \\
&-\dot{\beta}-\delta=-2\gamma \beta +4 \frac{\gamma k_B T}{m} \beta^2 \\
&-\dot{\delta}-2\alpha +2\omega^2 \beta=-\gamma \delta +4 \frac{\gamma k_B T}{m} \beta \delta,
\end{align}
\end{subequations}
with the following initial conditions
\begin{subequations}
\begin{align}
&\alpha(0)=\frac{m \omega_i^2}{2k_BT_i}, \quad \alpha(t_f)=\frac{m \omega_f^2}{2k_B T_f}\\
&\beta(0)=\frac{m}{2 k_B T_i}, \quad \beta(t_f)=\frac{m}{2 k_B T_f}\\
&\delta(0)=\delta(t_f)=0.
\end{align}
\end{subequations}

For the sake of simplicity, we now rescale all quantities and variables as follows
\begin{equation}
\widetilde{\alpha}\equiv \frac{2k_BT_i}{m \omega_i^2} \alpha, \qquad
 \widetilde{\beta}\equiv\frac{2k_BT_i}{m} \beta, \qquad
 \widetilde{\delta} \equiv \frac{k_BT_i}{m\omega_i} \delta, \qquad
\widetilde{\omega}\equiv \frac{\omega}{\omega_i}, \qquad
\widetilde{\kappa}\equiv \frac{\kappa}{\kappa_i}=\widetilde{\omega}^2, \qquad
 \widetilde{T}\equiv \frac{T}{T_i}, \qquad
 s=\frac{t}{t_f}.
\end{equation}
We distinguish the rescaled quantities from the corresponding dimensioned ones by a tilde. From now on, the dot stands for derivative with respect to rescaled time $s$. This rescaling yields the following set 
\begin{subequations}\label{set}
\begin{align}
&\dot{\widetilde{\alpha}}=2 N_{\omega} \widetilde{\kappa} \widetilde{\delta}-2N_{\gamma} \widetilde{T} \widetilde{\delta}^2 \\
&\dot{\widetilde{\beta}}=-2N_{\omega} \widetilde{\delta} +2N_{\gamma} \widetilde{\beta}-2 N_{\gamma} \widetilde{T} \widetilde{\beta}^2 \\
&\dot{\widetilde{\delta}}=- N_{\omega} \widetilde{\alpha} +  N_{\omega} \widetilde{\kappa} \widetilde{\beta}+N_{\gamma} \widetilde{\delta}-2 N_{\gamma} \widetilde{T} \widetilde{\beta} \widetilde{\delta} 
\end{align}		
\end{subequations}	
with initial conditions
\begin{subequations}
\begin{align}
&\widetilde{\alpha}(0)=1, \quad \widetilde{\alpha}(1)= \left(\frac{\omega_f}{\omega_i} \right)^2\frac{T_i}{T_f}\\
&\widetilde{\beta}(0)=1, \quad \widetilde{\beta}(1)=\frac{T_i}{T_f}\\
&\widetilde{\delta}(0)=\widetilde{\delta}(1)=0.
\end{align}
\end{subequations}
We introduced the two dimensionless parameters $N_{\gamma}=\gamma t_f$ and $N_{\omega}=\omega_i t_f$ that appear independently in Eqs.~\eqref{set}. They turn out to be two key parameters in the discussion within the underdamped framework. Indeed, most of the physics of the problem stems from the comparison between the characteristic timescales, namely the time associated with viscous friction 
$1/\gamma$, the period of oscillation in the harmonic potential $1/\omega$ and the duration $t_f$ of the protocol, or rather the position and velocity relaxation times $t_x=\gamma/\omega_i^2$ and $t_v=1/\gamma$, as will be discussed later on. The two numbers 
$N_{\gamma}$ and $N_{\omega}$ then simply compare the duration of the protocol with respectively the viscous time and the duration of an oscillation in the potential. 
Note that for clarity, we use the initial oscillation time to define $N_{\omega}$, as it would otherwise vary during the transformation, preventing us from performing a general analysis.

In an ESE approach, as highlighted above,  we choose the dynamics of some of the parameters of the particle density function (here among $\widetilde{\alpha}$, $\widetilde{\beta}$ and 
$\widetilde{\delta}$, or combinations of them) and deduce from them the required temporal evolution of the control parameters: here $\widetilde\kappa$ (and possibly also the bath temperature \cite{Chupeau:2018a}),
that can be controlled experimentally. 
The three relations~\eqref{eqESEresc} include five unknowns ($\widetilde{\alpha}$, $\widetilde{\beta}$, $\widetilde{\delta}$, $\widetilde{\kappa}$ and $\widetilde{T}$). Therefore two of the three functions 
$\widetilde{\alpha}$, $\widetilde{\beta}$ and $\widetilde{\delta}$ can be chosen freely, while the last function and the two control parameters are extracted from the equations. 
This choice is not unique, which is often an advantage, for it opens for a great versatility
of STA or ESE protocols; we will next see two particular ways to proceed.

\subsection{Specification for a fixed temperature bath}

The equations that control the system couple non linearly the functions $\widetilde{\alpha}$, $\widetilde{\beta}$ and $\widetilde{\delta}$ and the control parameters $\widetilde{\kappa}$ 
and $\widetilde{T}$, turning out to be rather complex to solve explicitly. As building an {\it explicit} protocol is desirable for the theoretical study of its properties as well 
as for the experimental implementation of the transformation, we further restrict our investigation to isothermal transformations $\widetilde{T}=1$. In this case, our set of three equations becomes
\begin{subequations}\label{eqESEresc}
\begin{align}
&\dot{\widetilde{\alpha}}=2 N_{\omega} \widetilde{\kappa} \widetilde{\delta}-2N_{\gamma} \widetilde{\delta}^2\label{alpha} \\
&\dot{\widetilde{\beta}}=-2N_{\omega} \widetilde{\delta} +2N_{\gamma} \widetilde{\beta}-2 N_{\gamma} \widetilde{\beta}^2 \label{beta}\\
&\dot{\widetilde{\delta}}=- N_{\omega} \widetilde{\alpha} +  N_{\omega} \widetilde{\kappa} \widetilde{\beta}+N_{\gamma} \widetilde{\delta}-2 N_{\gamma} \widetilde{\beta} \widetilde{\delta} \label{delta}
\end{align}		
\end{subequations}	
and the initial conditions are
\begin{subequations}
\begin{align}
&\widetilde{\alpha}(0)=1, \quad \widetilde{\alpha}(1)= \left(\frac{\omega_f}{\omega_i} \right)^2=\chi\\
&\widetilde{\beta}(0)= \widetilde{\beta}(1)=1 \\
&\widetilde{\delta}(0)=\widetilde{\delta}(1)=0
\end{align}
\label{eq:bc}
\end{subequations}
where $\chi=\kappa_f/\kappa_i$ is the compression factor. Although the bath temperature $\widetilde{T}$ is constant, the kinetic temperature $1/\widetilde{\beta}$ is in general not.
The overdamped limit of this isothermal case, where the kinetic temperature is supposed to stay at equilibrium at all times and consequently discarded from the treatment,
was previously addressed in~\cite{Martinez:2016b}. Here, in the underdamped regime, the problem is intrinsically more complex as both position and velocity distributions ({\it via} the functions $\widetilde{\alpha}$, $\widetilde{\beta}$ and $\widetilde{\delta}$) need to be engineered only through the stiffness $\widetilde{\kappa}$. 

In this particular isothermal case, we rephrase our set of three equations in a way that naturally leads to an explicit expression of the whole protocol. We first introduce the quantity $a$ and its rescaled equivalent $\widetilde{a}$
\begin{align}\label{defa}
&a= \alpha-\frac{\delta^2}{4\beta} \nonumber \\
&\widetilde{a}=\frac{2k_BT_i}{m\omega_i^2}a=\widetilde{\alpha}-\frac{\widetilde{\delta}^2}{\widetilde{\beta}}.
\end{align}
It directly relates to the position-variance (width of the marginal distribution of the position of the particle)
\begin{equation}\label{marginal}
P(x,t)=\int_{-\infty}^{+\infty} dv \, K(x,v,t)= \sqrt{\frac{a(t)}{\pi}} \exp(-a(t) x^2).
\end{equation}
Consequently $a$ is a measurable quantity, whereas the physical information borne by $\alpha$ is more elusive,
and it is advantageous to
work with $\widetilde{a}$ rather than $\widetilde{\alpha}$. Using 
equations~\eqref{eqESEresc}, it is easy to obtain a differential equation on $\widetilde{a}$
\begin{equation}\label{a}
\dot{\widetilde{a}} \,=\, 2 \, N_{\omega} \, \frac{\widetilde{\delta}}{\widetilde{\beta}} \, \widetilde{a}.
\end{equation}
Then we extract $\widetilde{\delta}$ from Eq.~\eqref{beta} and get
\begin{equation}\label{betaa}
\dot{\widetilde{\beta}}=-\widetilde{\beta} \frac{\dot{\widetilde{a}}}{\widetilde{a}} + 2N_{\gamma} \widetilde{\beta}-2 N_{\gamma} \widetilde{\beta}^2.
\end{equation}
It is convenient to define 
\begin{equation}\label{bigdeltadef}
\widetilde{\Delta}=\widetilde{a}\widetilde{\beta}
\end{equation}
in order to recast equation~\eqref{betaa} as
\begin{equation}\label{bigdelta}
\dot{\widetilde{\Delta}}=2N_{\gamma} \widetilde{\Delta} \left(1-\widetilde{\beta}\right).
\end{equation}
It turns out that all the used quantities can be expressed in terms of $\widetilde{\Delta}$ and its derivatives. Indeed, from equation~\eqref{bigdelta},  $\widetilde{\beta}$ reads
\begin{equation}
\widetilde{\beta}=1-\frac{\dot{\widetilde{\Delta}}}{2N_{\gamma} \widetilde{\Delta}} \label{betares}
\end{equation}
and $\widetilde{a}$ follows from the definition of $\widetilde{\Delta}$ (equation~\eqref{bigdeltadef}). Then the expression of $\widetilde{\delta}$ stems from equation~\eqref{a}
\begin{equation}\label{deltares}
\widetilde{\delta}=\frac{\widetilde{\beta} \dot{\widetilde{a}}}{2N_{\omega} \widetilde{a}},
\end{equation}
and $\widetilde{\alpha}$ from its definition~\eqref{alpha} together with equations~\eqref{betares} and~\eqref{deltares}. Finally, the control parameter $\widetilde{\omega}(s)$ can be read on 
equation~\eqref{alpha}
\begin{equation}\label{omega}
\widetilde{\kappa}\, =\, \frac{\dot{\widetilde{\alpha}}}{2N_{\omega}\widetilde{\delta}} \,+\, \frac{N_{\gamma}}{N_{\omega}} \,\widetilde{\delta}.
\end{equation}

\subsection{Towards explicit protocols}
\label{sec:explicit}

The above hierarchy of equations is convenient to devise explicit protocols, and they consequently fully qualify as 
STA or ESE in spirit. Indeed, the starting point provided by Eqs. (\ref{eqESEresc}) is not convenient, for these coupled equations involve the 
unknown forcing time-dependent term $\widetilde \kappa$, a function that needs to obey subtle properties to be compatible
with the boundary conditions \eqref{eq:bc}. Thus, as such, the problem is not amenable to numerical solution,
since it is of a functional-shooting type: find the proper family of $\widetilde\kappa(s)$ enforcing the desired 
final condition. On the other hand,  equations~\eqref{bigdeltadef},~\eqref{betares},~\eqref{deltares},~\eqref{alpha} and~\eqref{omega}
pave the way to a simple analytical solution:
we first choose the shape of $\widetilde{\Delta}(s)$ and subsequently deduce from it the evolution of $\widetilde \beta$ making use of \eqref{betares}, from which $\widetilde a$ is known invoking 
\eqref{bigdeltadef}; $\widetilde \delta$ then follows from \eqref{deltares}, $\widetilde \alpha$ from \eqref{alpha} and the desired
forcing is computed, at the end of the chain, with Eq. \eqref{omega}. That solution is referred to as Protocol A.

We emphasize that special attention has to be paid to the temporal boundary conditions, as in every shortcut-to-adiabaticity-type procedure, in order to avoid excitations of the system when 
reaching equilibrium, at the end of the protocol. In addition,
a protocol that is smooth enough at initial time, when launched, is more conducive to a successful experimental realization.
We thus enforce the same boundary conditions at initial time $s=0$ and final time $s=1$, but it can be kept in mind that a different choice
can be made at $s=0$, under the proviso that the boundary conditions at $s=1$ are as above. Equations~\eqref{eqESEresc} imply that the first derivative of $\widetilde{\alpha}$, $\widetilde{\beta}$ 
and $\widetilde{\delta}$ vanishes at initial and final times, and so does the first derivative of $\widetilde{a}$. This condition on the first derivative of $\widetilde{\delta}$ also 
forces the second derivative of $\widetilde{a}$ to be zero, through equation~\eqref{deltares}. In turn, this imposes the same condition on the second derivative of $\widetilde{\beta}$. 
Altogether, this requires at least that the first three derivatives of $\widetilde{\Delta}(s)$ are zero at the final time. 
Finally, the values of $\widetilde{\Delta}$ at initial 
and final times are simply $\widetilde{\Delta}(0)=1$ and $\widetilde{\Delta}(1)=\chi$. For the sake of simplicity, we choose a polynomial for $\widetilde{\Delta}$. 
The lowest order admissible function reads
\begin{equation}\label{Delta}
\widetilde{\Delta}(s)=1+ \left(\chi-1\right) \left(35s^4-84s^5+70s^6-20s^7\right).
\end{equation}
This method is straightforward, and singles out $\widetilde\Delta$, over which ``control'' is exerted.
It is a combination between the width of the position distribution 
and the kinetic temperature (related to the velocity distribution), and thus a rather ``secondary'' quantity.
Our goal is next to present an explicit variant, where another quantity is controlled, with more
direct physical meaning. \\

Another route towards an explicit protocol consists in choosing the inverse kinetic temperature $\widetilde{\beta}(s)$ and deducing from it the other functions, 
through the same aforementioned hierarchy. The temporal boundary conditions required for $\widetilde{\beta}(s)$ are again $\widetilde{\beta}(0)=\widetilde{\beta}(1)=1$ 
and the first two derivatives vanish at initial and final times. Once $\widetilde{\beta}(s)$ is chosen, 
$\widetilde{a}(s)$ is obtained integrating equation~\eqref{betaa}
\begin{equation}\label{asol}
\widetilde{a}(s)=\frac{1}{\widetilde{\beta}(s)} \exp\left( 2 N_{\gamma} \int_0^s du  \left( 1-\widetilde{\beta}(u) \right)  \right)
\end{equation}
while the other quantities can be expressed in terms of functions $\widetilde{a}$ and $\widetilde{\beta}$. The initial condition $\widetilde{a}(0)=1$ is fulfilled since $\widetilde{\beta}(0)=1$, but the final condition $\widetilde{a}(1)=\chi$ imposes an additional integral constraint on the chosen $\widetilde{\beta}(s)$
\begin{equation}\label{constraint}
2 N_{\gamma} \int_0^1 ds  \left( 1-\widetilde{\beta}(s) \right)  = \ln \chi.
\end{equation}
For the previous procedure, only $\widetilde{\Delta}$ and its derivatives were employed to express all the other functions, so that specifying the boundary conditions of 
$\widetilde{\Delta}$ was enough to ensure the right initial and final states. Here, the solving procedure involves an integral of the chosen function $\widetilde{\beta}(s)$ 
in equation~\eqref{asol}. As a result, the initial and final states cannot be both encoded in the temporal boundary condition of $\widetilde{\beta}$ and yield the 
integral constraint~\eqref{constraint}. Note that this relation is true whatever the protocol, and is in particular automatically verified for Protocol A when $\widetilde{\Delta}$ has smooth enough boundary conditions.
We coin this variant ``Protocol B''.

Finally, a more natural quantity to choose and control would be the inverse variance of position $\widetilde{a}(s)$. However, there is no straightforward solution of the equations 
in terms of $\widetilde{a}(s)$, as there was in terms of $\widetilde{\Delta}$ or $\widetilde{\beta}$. Moreover, a numerical resolution of the hierarchy of equations in 
terms of $\widetilde{a}$ necessarily involves a numerical integration of a differential equation, for example equation~\eqref{asol}. 
Even a careful choice of the boundary conditions of $\widetilde{a}$ fails to produce reliably the desired initial and final states for the transformation. 
We meet again a functional shooting problem, where the function $\widetilde{a}(s)$, which is a parameter of the following equation 
\begin{equation}
\dot{\widetilde{\Delta}}=2 N_{\gamma} \widetilde{\Delta} \left( 1- \frac{\widetilde{\Delta}}{\widetilde{a}} \right),
\end{equation}
has to be tuned so that the solution $\widetilde{\Delta}$ fulfills the desired initial and final conditions. 
Equipped with protocols A and B, we are nevertheless in a position to discuss the robustness of our main findings.

%%%%%%%%%%%%%%%%%%%%%%%%%%%%%%%%%%%%%%%%%%%%%%%%%%%%%%%%%%%%%%%%%%%%%%%%%%%%%%%%
\section{Charting out the phase diagrams of the problem}
\label{sec4}

\begin{figure}[h!]
\centering
\includegraphics[width=0.7\columnwidth]{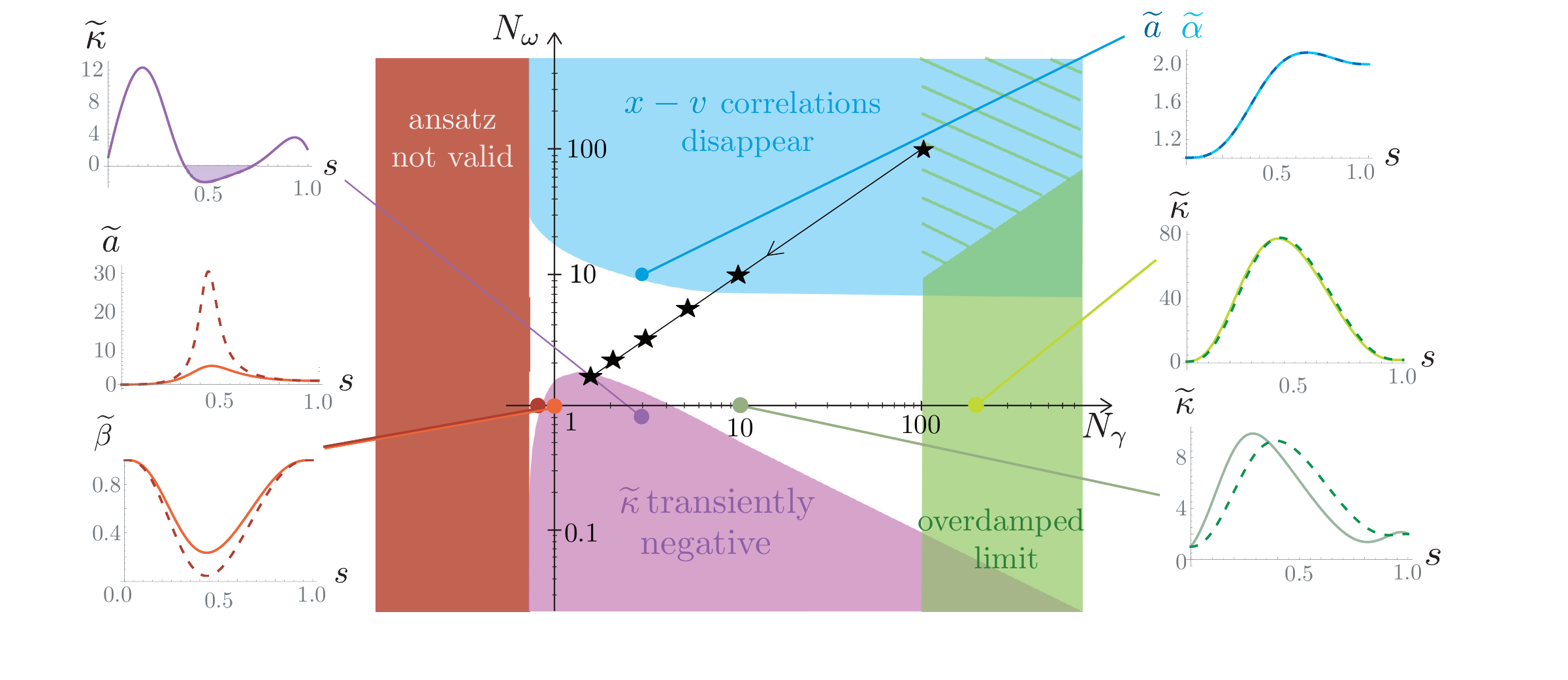}
\caption{Phase diagram of the underdamped protocol A for a compression in the ($N_{\gamma}\equiv \gamma t_f,N_{\omega}\equiv \omega_i t_f$) plane, in a log-log scale. 
The colors of the inset curves correspond to the different sectors defined (while the white region is ``neutral'' regarding all the properties discussed in this section). The overdamped regime is represented in green and can be extended to the hatched region (see paragraph~\ref{sec:over}). In the middle and bottom right insets,  
the solid curves display the stiffness $\widetilde{\kappa}$ (computed within the full underdamped formalism),
while the dashed curves are for its overdamped counterpart. The middle right inset is for a state point within the 
overdamped region, while the bottom right is not (and hence, the dashed and continuous curves are distinct). 
The $x-v$ decoupling zone is blue (see paragraph~\ref{sec:decorr}), 
and the corresponding inset shows the collapsed evolution of $\widetilde{a}$ and $\widetilde{\alpha}$. The zone of non-existence of the protocol is represented in red (see paragraph~\ref{sec:exist}), 
and the behavior of the functions $\widetilde{a}$ and $\widetilde{\beta}$ near the boundary of this region is shown in the left inset in red. 
Finally, the region where $\widetilde{\kappa}$ becomes transiently negative (see paragraph~\ref{sec:pbs}) is in purple. Here, the compression factor is $\chi=2$, \smash{$\widetilde{\Delta}$} is given by equation~\eqref{Delta}, and the set of ($N_{\gamma},N_{\omega}$) parameters is 
indicated by the position of the dots on the phase diagram. The straight thin black line represents a ``trajectory'' followed in the phase diagram when the duration $t_f$ of the protocol is decreased, all other parameters being fixed (see Fig.~\ref{traj} for further details). 
The stars correspond to the different curves presented in Figure~\ref{traj}.
}
\label{phasediag}
\end{figure}

We now study the characteristics of protocol A, as a function of the two dimensionless parameters $N_{\gamma}$ and $N_{\omega}$. We focus 
in particular on the very existence of the protocol, on the 
applicability of the overdamped limit, on the decoupling of the $x$ and $v$ degrees of freedom, and on the temporal evolution of the only control 
parameter of the problem, namely the stiffness $\widetilde{\kappa}$ of the trap. We represent this thorough study on two phase diagrams in Figures~\ref{phasediag} and~\ref{phasediag2}. 
They of course depend on the characteristics of the desired transformation, as will be discussed later on; we have chosen here a compression factor $\chi=2$ for Figure~\ref{phasediag} and a decompression 
factor $\chi=0.5$ for Figure~\ref{phasediag2}.

\subsection{Zones of the phase diagrams}

\subsubsection{Existence of the protocol}\label{sec:exist}

We first investigate the existence of the underdamped protocol itself. For the underdamped ansatz~\eqref{ansatz} to be well-defined, the functions $\widetilde{\alpha}$ 
and $\widetilde{\beta}$, as well as $\widetilde{a}$, must remain positive during the whole transformation; our ansatz is otherwise divergent.
However, the expression~\eqref{betares} shows that $\widetilde{\beta}$  is positive only as long as
\begin{equation}
\dot{\widetilde{\Delta}}<2 N_{\gamma} \widetilde{\Delta}.
\end{equation}
This criterion can always be satisfied in the case of a decompression, where $\widetilde{\Delta}$ can be chosen monotonically decreasing whatever the value of $N_{\gamma}$, 
while $\widetilde{\Delta}$ is by construction always positive. On the other hand, for a compression, low values of $N_{\gamma}$ 
can make the function $\widetilde{\beta}$ become negative, resulting in a diverging ansatz. This happens typically when $N_{\gamma}$ is of order 1. 
The exact threshold of course depends on the chosen shape of $\widetilde{\Delta}$ and the compression factor $\chi$, {as will be discussed later}. Moreover, since the product 
$\widetilde{\Delta}=\widetilde{a} \widetilde{\beta}$ is fixed and positive, $\widetilde{a}$ and $\widetilde{\beta}$ are always of the same sign. As for $\widetilde{\alpha}$, 
it is also positive when $\widetilde{\beta}$ (and thus $\widetilde{a}$) is positive too. 
Therefore, the only region of the compression phase diagram where the ansatz is ill-defined is the left half-plane under a threshold $N^{\rm min}_{\gamma}$ of order unity. 
The ansatz is on the other hand well-defined in the whole decompression phase diagram.

We illustrate this on Figure~\ref{phasediag}, where $\widetilde{a}$ and $\widetilde{\beta}$ are plotted near the boundary of the existence region, on the red left inset curves. 
We notice that the minimum of $\widetilde{\beta}$ tends quite quickly to zero when $N_{\gamma}$ is decreased around one and that in turn, $\widetilde{a}$ becomes very large in this region. 
When the inverse kinetic temperature $\widetilde{\beta}$  goes to zero, the velocity distribution of the particle 
becomes very broad and the kinetic temperature diverges. In order to keep a control on the particle, the stiffness of 
the confinement has to increase dramatically, yielding a very peaked $\widetilde{a}$, as illustrated.
In this region, where the protocol can be considered fast since it no longer allows for velocity equilibration,
it is necessary to use large stiffness to obtain the desired compression, which provides work to the system. 
This results in a heating, marked by the drop of $\beta$. 

Our approach fails when the duration $t_f$ of the compression protocol is of the order of the friction time $1/\gamma$ (see Figure~\ref{phasediag}, left area). This can be overcome by using a non-conservative potential that would penalize high velocities and then prevent this dramatic increase of the kinetic temperature, as suggested in~\cite{Tu:2014,Li:2017}. On the other hand, for a decompression, the kinetic temperature cannot diverge during the transformation, and our 
approach remains valid at any friction in this case, without needing a non-conservative potential. We stress that realizing such forces in an experiment is a challenge,
while the conservative case worked out here is routinely employed with optically confined colloids (see \cite{ber14} and references therein).

\subsubsection{Overdamped limit}\label{sec:over}

Next, we consider the overdamped limit and show that we can recover the results obtained in~\cite{Martinez:2016b} from the underdamped formalism developed here. We also determine the 
regime of validity of the overdamped approximation, which amounts to treating the velocity degrees of freedom as equilibrated at all times. As their distribution relaxes to the 
Gaussian distribution on a timescale $1/\gamma$, the overdamped approximation requires this time to be much smaller than the other timescales of the problem;  
in our notations, this leads to $N_{\gamma} \gg 1$ and 
$N_{\gamma} \gg N_{\omega}$. Note that in the following, $\widetilde{\Delta}(s)$ is kept unspecified for the sake of generality, but is always of ``order 1'', in that it is chosen 
independently of $N_{\gamma}$ and $N_{\omega}$. Moreover, the following discussion holds both for a compression and a decompression. 

\begin{figure}[h!]
\centering
\includegraphics[width=0.8\columnwidth]{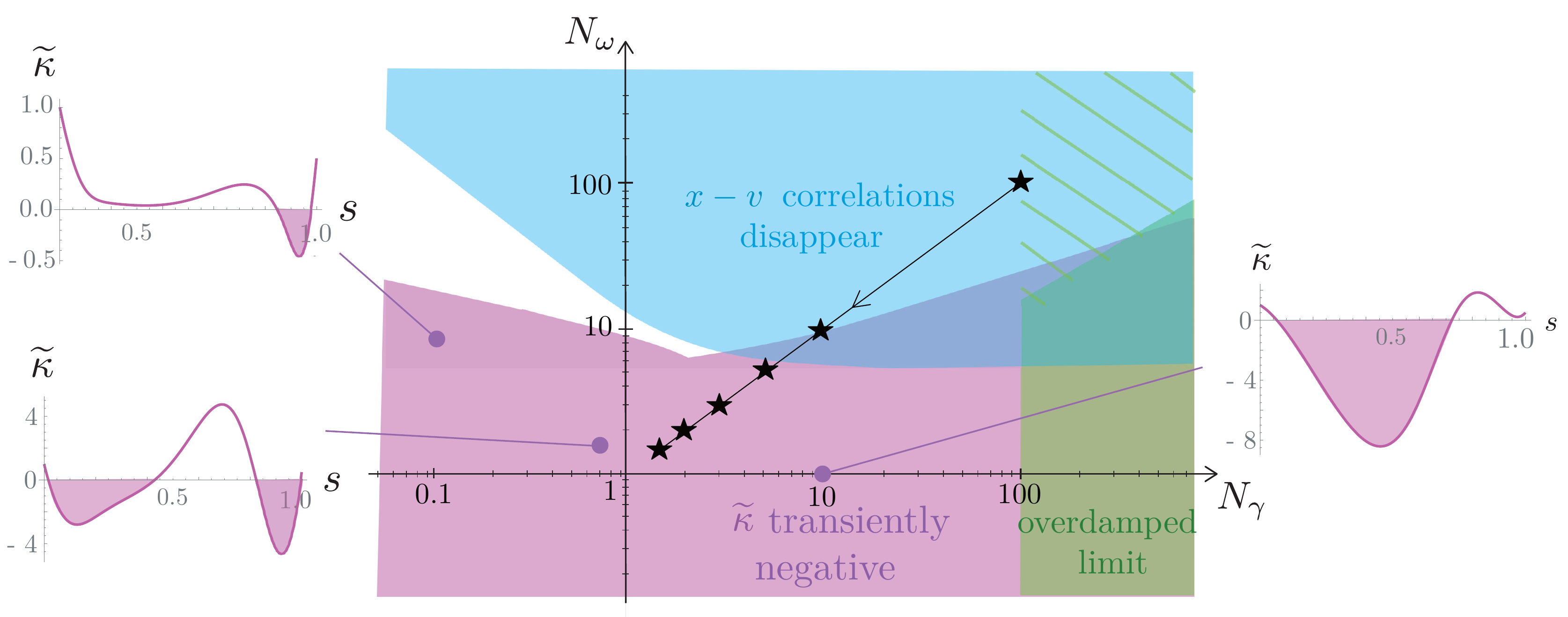}
\caption{Phase diagram of the underdamped protocol A for a decompression in the ($N_{\gamma},N_{\omega}$) plane.  The color code is the same as in Figure~\ref{phasediag}, 
and the compression factor $\chi$ is 0.5. In a symmetric fashion compared to quick compressions where the stiffness exhibits strong overshoots, quick decompressions rely on a stiffness that involves a strong undershoot (see the three insets). Exceeding the target value of the stiffness (upwards for a compression, downwards for a decompression) is a means to accelerate the transformation. For very quick transformations, the shape of the stiffness can become very complex and display two undershoots as illustrated by the left two insets. Again, the straight thin black line represents a ``trajectory'' followed in the phase diagram when the duration $t_f$ of the protocol is decreased, all other parameters being fixed, 
and the stars correspond to the different curves of Fig.~\ref{traj}.
}
\label{phasediag2}
\end{figure}

To investigate the overdamped limit, we focus on the temporal evolution of the stiffness of the potential $\widetilde{\kappa}$.
Starting from the overdamped formalism, it was found in~\cite{Martinez:2016b} to be related to the function $\widetilde{a}(s)$ of equation~\eqref{marginal} through
\begin{equation}\label{overresc}
\widetilde{\kappa}-\widetilde{a} = \frac{N_{\gamma}}{2N_{\omega}^2} \frac{\dot{\widetilde{a}}}{\widetilde{a}}.
\end{equation}
Therefore, we concentrate on the quantity $\widetilde{\kappa}-\widetilde{a}$ in the regime $N_{\gamma} \gg 1$ and $N_{\gamma} \gg N_{\omega}$. 
Using equation~\eqref{alpha},~\eqref{defa} and~\eqref{a}, we obtain
\begin{equation}\label{under}
 \widetilde{\kappa}-\widetilde{a}= \widetilde{a}\left( \frac{1}{\widetilde{\beta}}-1\right) +\frac{ \widetilde{a}}{\widetilde{\beta}\dot{\widetilde{a}}} 
\left(  2\frac{\widetilde{\delta} \dot{\widetilde{\delta}}}{ \widetilde{\beta}} - \frac{\widetilde{\delta}^2 \dot{\widetilde{\beta}}}{ \widetilde{\beta}^2} \right) +  
\frac{N_{\gamma} \dot{\widetilde{a}}}{2N_{\omega}^2 \widetilde{a}}  \widetilde{\beta}.
\end{equation}
This expression can be developed in power series of $N_{\gamma}$. We emphasize that in the limit of large $N_{\gamma}$, $\widetilde{\beta}$ 
remains close to one (as expected, the kinetic temperature is equilibrated at all times with the bath temperature), and $\widetilde{a}$ is then approximately $\widetilde{\Delta}$, 
which does neither depend on $N_{\gamma}$ nor on $N_{\omega}$. This yields
\begin{equation}\label{part1}
 \widetilde{a}\left( \frac{1}{\widetilde{\beta}}-1\right)=\frac{\widetilde{a} \dot{\widetilde{\Delta}}}{4 N_{\gamma} \widetilde{\Delta}}+ {\cal O}\left( \frac{1}{N_{\gamma}^2}\right)
\end{equation}
and
\begin{equation}\label{part2}
\frac{ \widetilde{a}}{\widetilde{\beta}\dot{\widetilde{a}}} \left(  2\frac{\widetilde{\delta} \dot{\widetilde{\delta}}}{ \widetilde{\beta}} - \frac{\widetilde{\delta}^2 \dot{\widetilde{\beta}}}{ \widetilde{\beta}^2} \right) +  \frac{N_{\gamma} \dot{\widetilde{a}}}{2 N_{\omega}^2 \widetilde{a}}  
\widetilde{\beta}=\frac{N_{\gamma} \dot{\widetilde{a}}}{2 N_{\omega}^2 \widetilde{a}}  \widetilde{\beta}\left[ 1 +{\cal O}\left( \frac{1}{N_{\gamma}} \right) \right].
\end{equation}
Finally under the assumption $N_{\gamma} \gg N_{\omega}$, we can compare the leading term of equations~\eqref{part1} and~\eqref{part2}
\begin{equation}
\left| \frac{\widetilde{a} \dot{\widetilde{\Delta}}}{2 N_{\gamma} \widetilde{\Delta}} \right|  \ll  \left| \frac{N_{\gamma} \dot{\widetilde{a}}}{N_{\omega}^2 \widetilde{a}}  \widetilde{\beta} \right|,
\end{equation}
and we obtain
\begin{equation}\label{dvlpt}
 \widetilde{\kappa}- \widetilde{a}=\frac{N_{\gamma}}{2N_{\omega}^2} \frac{\dot{\widetilde{a}}}{\widetilde{a}} \left[ 1+ {\cal O}\left(\frac{1}{N_{\gamma}} \right) + 
{\cal O}\left( \frac{N_{\omega}^2}{N_{\gamma}^2} \right) \right].
\end{equation}
This matches exactly equation~\eqref{overresc} in the overdamped limit, as expected when $N_{\gamma} \gg N_{\omega}$ and $N_{\gamma} \gg 1$. This expression is also informative on the scalings with respect to $N_\gamma$ and $N_\omega$ of the corrections to the 
overdamped asymptotics. Yet, a subtlety should be outlined here, and it is related to the polynomial shape chosen 
for $\widetilde \Delta(s)$. When our parameters qualify the protocol as belonging in the overdamped regime, $\widetilde a$ inherits this
functional form, which is slightly more complex than the one chosen in \cite{Martinez:2016b}, where the smoothness requirement only concerns its first derivative. Indeed, the underdamped protocol requires more derivatives to vanish than its overdamped counterpart as discussed in paragraph~\ref{sec:explicit}, which results 
in a polynomial $\widetilde \Delta(s)$ of higher order. 
Note here that comparing directly the stiffness $\widetilde{\kappa}$ computed within the full underdamped formalism to the overdamped one, instead of focusing on $\widetilde \kappa-\widetilde a$, yields a looser criterion: 
$N_{\gamma} \gg 1$ without any condition on $N_{\omega}$, as represented on the phase diagrams~\ref{phasediag} and~\ref{phasediag2} in hatched green.

To conclude this discussion, we point out that our criterion for the overdamped limit differs from the one given in~\cite{Li:2017}, which can be rephrased as $N_{\omega} \ll N_{\gamma}$. 
This discrepancy likely comes from the fact that the authors impose that the temperature of the particle (our inverse $\beta$) remains constant during the transformation. 
Though this choice offers much simpler calculations, it requires to implement a potential that is quadratic in impulsion, therefore creating a set-up that strongly strays from ours, 
where the protocol only involves conservative drivings.

\subsubsection{$x-v$ decorrelation}\label{sec:decorr}

An interesting  characteristics of the protocol is whether or not position and velocity degrees of freedom are correlated. This correlation is measured by comparing 
the (squared) cross term $\widetilde{\delta}^2$ to the product of position and velocity variances
$\widetilde{\alpha} \widetilde{\beta}$. In particular, these correlations vanish when $\widetilde{a}$ collapses onto $\widetilde{\alpha}$. 
Following the definition of $\widetilde{a}$ in \eqref{defa} and of $\widetilde{\Delta}$ in~\eqref{bigdeltadef}, this amounts to the criterion $|\widetilde{\delta} |^2 \ll |\widetilde{\Delta}|$. 

Going back to  expression~\eqref{deltares} for $\widetilde{\delta}$, we see that its scaling depends on that of $\widetilde{\beta}$. At large $N_{\gamma}$, 
$\widetilde{\beta} \simeq 1$ so that $\widetilde{\delta}$ scales as $1/N_{\omega}$. On the contrary, for low $N_{\gamma}$ (that only concern decompressions as discussed 
in the paragraph about the existence of the protocol), $\widetilde{\beta}$ scales as $1/N_{\gamma}$ so that $\widetilde{\delta}$ scales as $1/(N_{\omega} N_{\gamma})$. The criterion 
$|\widetilde{\delta}|^2 \ll |\widetilde{\Delta}|$ then takes two different shapes in the limits of high and low $N_{\gamma}$. For high $N_{\gamma}$, $x$ and $v$ degrees of 
freedom are decorrelated for $N_{\omega}^2 \gg 1$, which is confirmed by the two phase diagrams (see Figures~\ref{phasediag} and~\ref{phasediag2}). 
On the other hand, for low $N_{\gamma}$ and for decompressions, decorrelation arises when $N_{\omega} \gg 1/N_{\gamma}$. This is also confirmed by the decompression phase diagram 
(Figure~\ref{phasediag2}), where the left frontier of the decorrelation area has a slope $-1$ in double log scale. 
The border of this region is determined numerically as the curve on which the relative difference between $\widetilde{a}$ and $\widetilde{\alpha}$ is of one percent.

\subsubsection{Implementation challenges and ESE relevance}\label{sec:pbs}

Finally, a primordial aspect of a protocol such as devised here lies in the characteristics of the control parameter $\widetilde{\kappa}$ that needs to be enforced experimentally to achieve the desired transformation. The main experimental challenge arises when the stiffness of the trap becomes transiently negative. In this case, the potential switches from confining to repulsive, a feature that cannot be achieved simply with an optically trapped colloid. We determine numerically the part of the phase diagram where this change in the sign of the trap curvature takes place; it is represented in purple on Figs.~\ref{phasediag} and~\ref{phasediag2}.

\begin{figure}[h!]
\centering
\includegraphics[width=480pt]{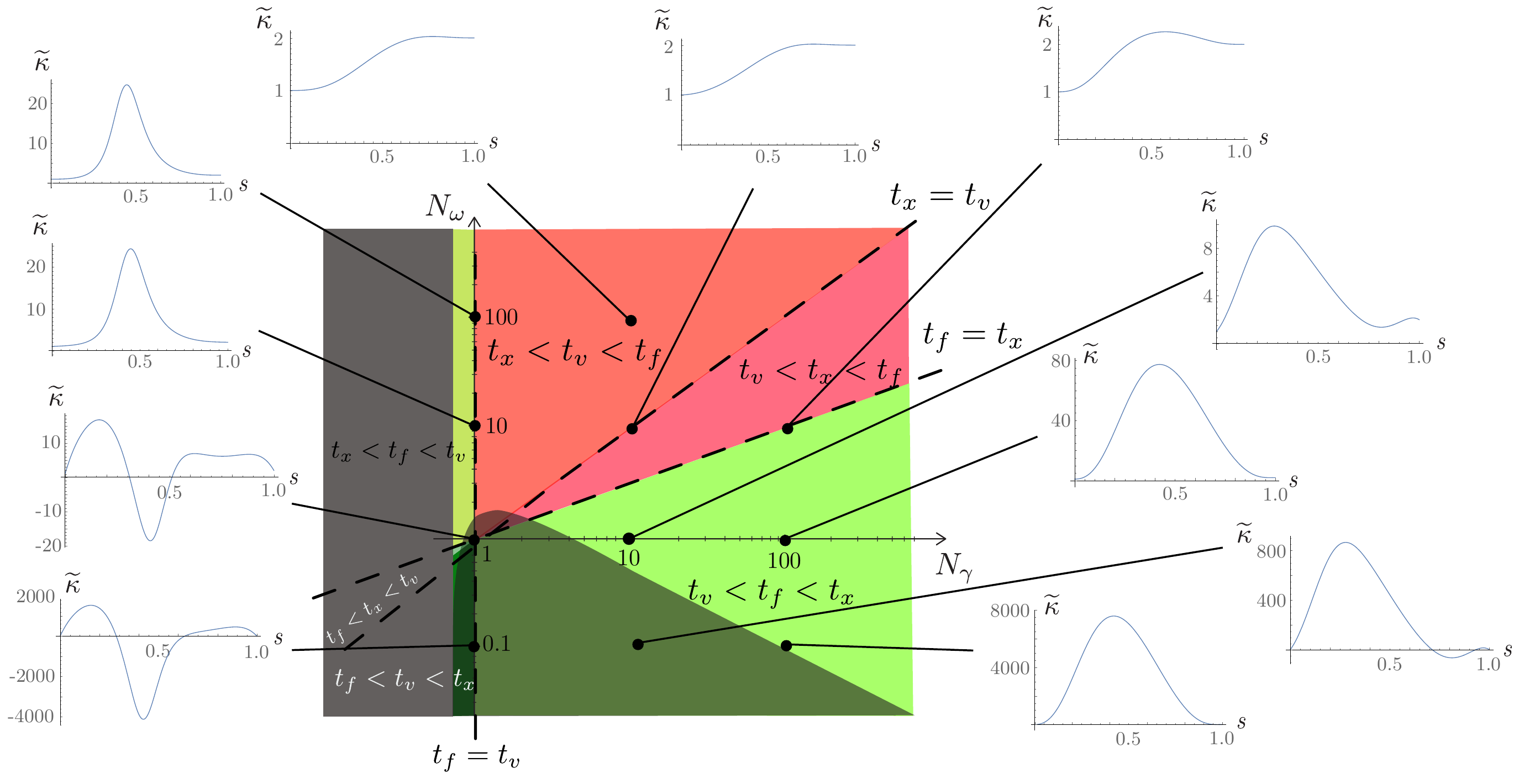} \\
\vspace*{0.5cm}
\caption{Compression (top) and decompression (bottom) diagrams in the plane $(N_{\gamma},N_{\omega})$, showing the behavior of the stiffness during the protocol, while comparing the three timescales, namely the duration of the protocol $t_f$, the position relaxation time $t_x$ and the velocity relaxation time $t_v$. Dashed lines indicate where two of these timescales are equal, and define six zones. The red ones represent the set of $(N_{\gamma},N_{\omega})$ for which the ESE protocol is slow ($t_f$ is the largest timescale) and then less interesting. The green ones on the contrary indicate areas where the ESE protocol is interesting for a relaxation shortcut ($t_f$ is not the largest timescale). The grey areas show features from Figs.~\ref{phasediag} and~\ref{phasediag2}, that is the non-existence zone for compressions and the negative stiffness zone.}
\label{diag2}
\end{figure}

The global behavior of the stiffness has a complex dependence on the physical parameters $N_{\gamma}$ and $N_{\omega}$. We capture this diversity with Figure~\ref{diag2}, where we represent the shape of the stiffness as a function of rescaled time $s$ for representative points of the phase space $(N_{\gamma},N_{\omega})$. From the phase diagrams of Figs.~\ref{phasediag} and~\ref{phasediag2}, we only keep the features that concern the implementation of the protocol, namely the non-existence area for compressions, and the aforementioned zone where the stiffness is transiently negative. Insights into the zoology of behaviors of the stiffness can be gained by comparing the three timescales of the problem, namely the duration of the protocol $t_f$, the timescale of position relaxation $t_x=\gamma/\omega_i^2$ and the timescale of the velocity relaxation $t_v=1/\gamma$, which are combined in $N_{\gamma}$ and $N_{\omega}$. Note that the maximum of these last two timescales defines the global relaxation timescale of the problem. On Figure~\ref{diag2}, the dashed lines indicate where these timescales are equal two by two, and the different colors stand for the six possible orders of the three timescales. We emphasize that for ESE purposes, the two red areas, where the protocol is slower than the global relaxation scale of the problem, are less interesting than the green areas where position or velocity (or both) relaxes more slowly than the protocol. The most interesting set of parameters $(N_{\gamma},N_{\omega})$ ESE-wise are therefore the green zones of these two diagrams out of the grey-shadowed areas.

Determining the shape of the stiffness from general considerations on our explicit protocol is challenging, as it involves several quantities that themselves depend non trivially on $N_{\gamma}$ and $N_{\omega}$ (see Eq.~\eqref{omega}). However, we can grasp some features of the shape of the stiffness with the help of the diagrams of Fig~\ref{diag2}. First, the stiffness has a very smooth behavior in the red zones, as expected, where the protocol is slow compared to the intrinsic dynamics of the system. At large $N_{\gamma}$, in the region where the stiffness is appropriately described by its overdamped expression~\eqref{dvlpt} as discussed in section~\ref{sec:over}, the dependence on the parameters $N_{\gamma}$ and $N_{\omega}$ of the stiffness required to perform the transformation is proportional to $N_{\gamma}/N_{\omega}^2$ (see Eq.~\eqref{overresc}), {\it i.e.} $t_x/t_f$. Then the region where the amplitude of the stiffness is large compared to 1 corresponds to the part of the diagrams that is well below the line $t_f=t_x$, as shown on the bottom right insets for compression diagram and most of the bottom insets for the decompression diagram. The farther from the line $t_f=t_x$, the greater the amplitude of the stiffness. This tendency to have large variations in the stiffness is also amplified when $t_f$ becomes smaller than $t_v$, as shown on the two bottom left insets of the decompression diagram. Finally, the regions in dark green, where ESE protocols are highly desirable because shorter than both position and velocity relaxation times, display extreme shapes of stiffness. These bizarre shapes result from the difficulty to control the evolution of both position and velocity, which cannot equilibrate themselves that quickly, with only one control parameter.

\subsection{Robustness of the phase diagrams}

We have analyzed above protocol A phase diagrams, in which some details can depend of the 
functional choice made for the chosen $\widetilde{\Delta}(s)$. We now quickly address protocol B 
properties, as a means to put to the test the robustness of our findings. In this discussion, 
we choose to only address protocols where the kinetic temperature is temporarily increased in the case of a compression (resp. decreased in the case of a decompression). 
In other words, we suppose that \smash{$\widetilde{\beta}(s)$} is always smaller than one (resp. bigger than one). This is equivalent to restricting to monotonous \smash{$\widetilde{\Delta}(s)$}, 
as indicated by equation~\eqref{bigdelta}. As $\widetilde{\Delta}(0)=1$ and $\widetilde{\Delta}(1)=\chi$, this function remains bounded, irrespective 
of the values of the parameters $N_{\gamma}$ and $N_{\omega}$. The previous discussions about the overdamped limit and the decorrelation of 
position and velocity degrees of freedom are therefore still valid.
These areas are then completely 
robust with respect to the exact shape of the protocol.

On the other hand, the exact position of the region where the ansatz no longer exists depends on the shape of the chosen function $\widetilde{\beta}$ for protocol B
or $\widetilde{\Delta}$ for protocol A. The corresponding boundary is always a vertical line in the compression phase diagram, as equation~\eqref{bigdelta} does not involve $N_{\omega}$. The most favorable case, 
in which the existence domain is the largest, corresponds to a situation where $\widetilde{\beta}(s)$ is flat during almost the whole compression (except near initial and 
final times to fulfill the temporal boundary conditions). Together with the fixed integral~\eqref{constraint} for $\widetilde{\beta}$, this indicates that the protocol-dependent existence 
threshold $N_{\gamma}^{\rm min}$ is always such that
\begin{equation}
N_{\gamma}^{\rm min} > \frac{\ln \chi}{2}.
\end{equation} 
Therefore this non-existence zone is a generic feature of the process and cannot be eliminated by tinkering with the shape of the protocol. In particular, there is no way to devise a very fast isothermal compression protocol:
$t_f > \gamma^{-1} (\ln\chi)/2$. This is a strict lower bound, that can be significantly exceeded in cases where $\widetilde{\Delta}$ is non-monotonous (protocol A), or equivalently when 
$\widetilde{\beta}$ presents an overshoot (protocol B). The integral of $\widetilde{\beta}(s)$ being fixed by equation~\eqref{constraint}, such an overshoot necessarily needs to 
be compensated by lower values of $\widetilde{\beta}(s)$ in a different part of the transformation, therefore approaching the forbidden zero value. 
This may endanger the convergence of the ansatz and thus provide a further reason to dismiss non-monotonous $\widetilde{\Delta}$.

\begin{figure}[h!]
\centering
\includegraphics[width=350pt]{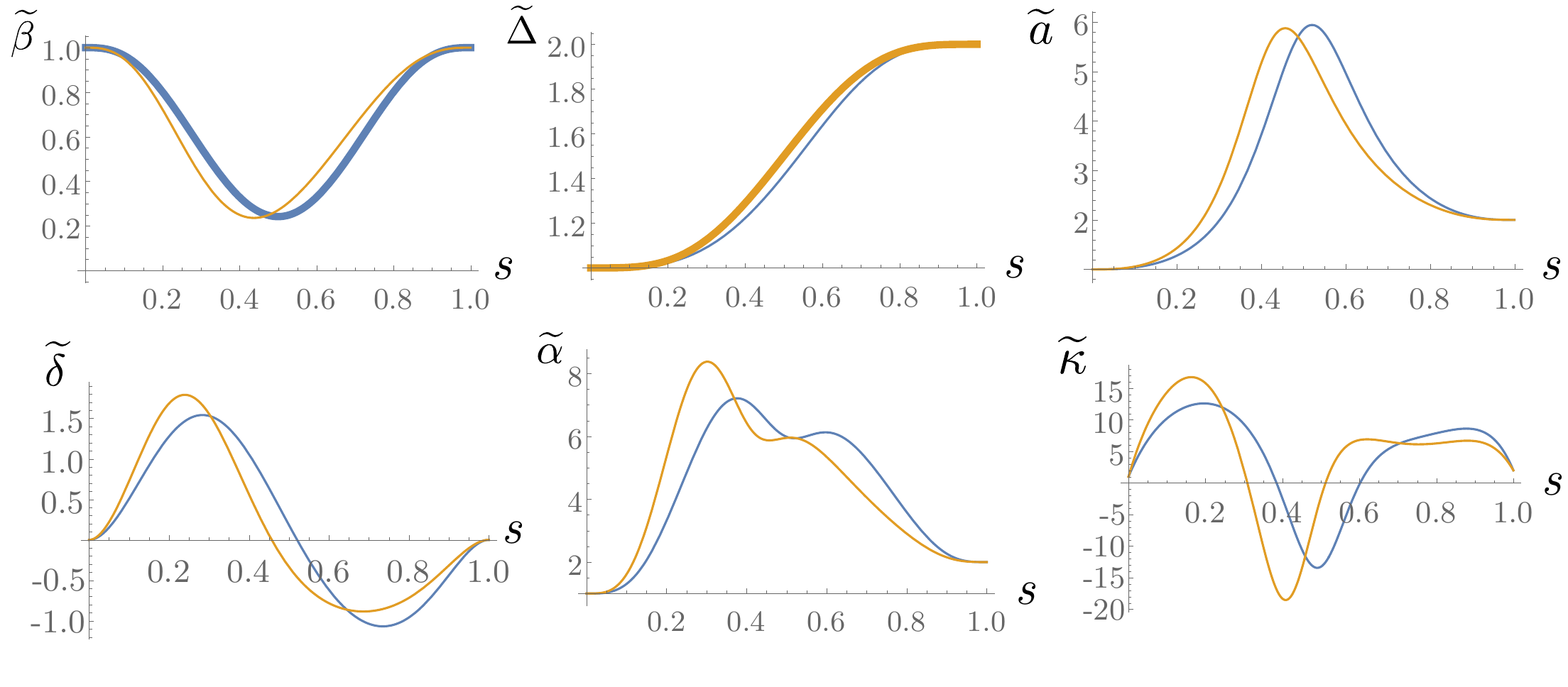}
\caption{Comparison between protocols A and B. The thicker curve corresponds to the chosen 
function from which the whole protocol is deduced, $\widetilde{\beta}=1-140 \ln \chi/(2 N_{\gamma}) s^3 (1-s)^3$ for protocol B (blue curves), and $\widetilde{\Delta}$ given 
in equation~\eqref{Delta} for protocol A (orange curves). The parameters are $\chi=2$, $N_{\gamma}=N_{\omega}=1$ (fast compression).}
\label{robustness}
\end{figure}
As for the rather exotic shapes of the stiffness $\widetilde{\kappa}$ discussed earlier, we show in Figure~\ref{robustness} 
the differences between protocols A and B. As expected from the hierarchy of equations, a slight modification in the driving 
function yields significant changes, affecting all other quantities, including $\widetilde{\kappa}$.
However, Fig.~\ref{robustness} indicates that the essential features reported in the phase diagrams are robust with respect to a protocol change.

\subsection{Consequences of accelerating the protocol}

\begin{figure}[h!]
\centering
\includegraphics[width=400pt]{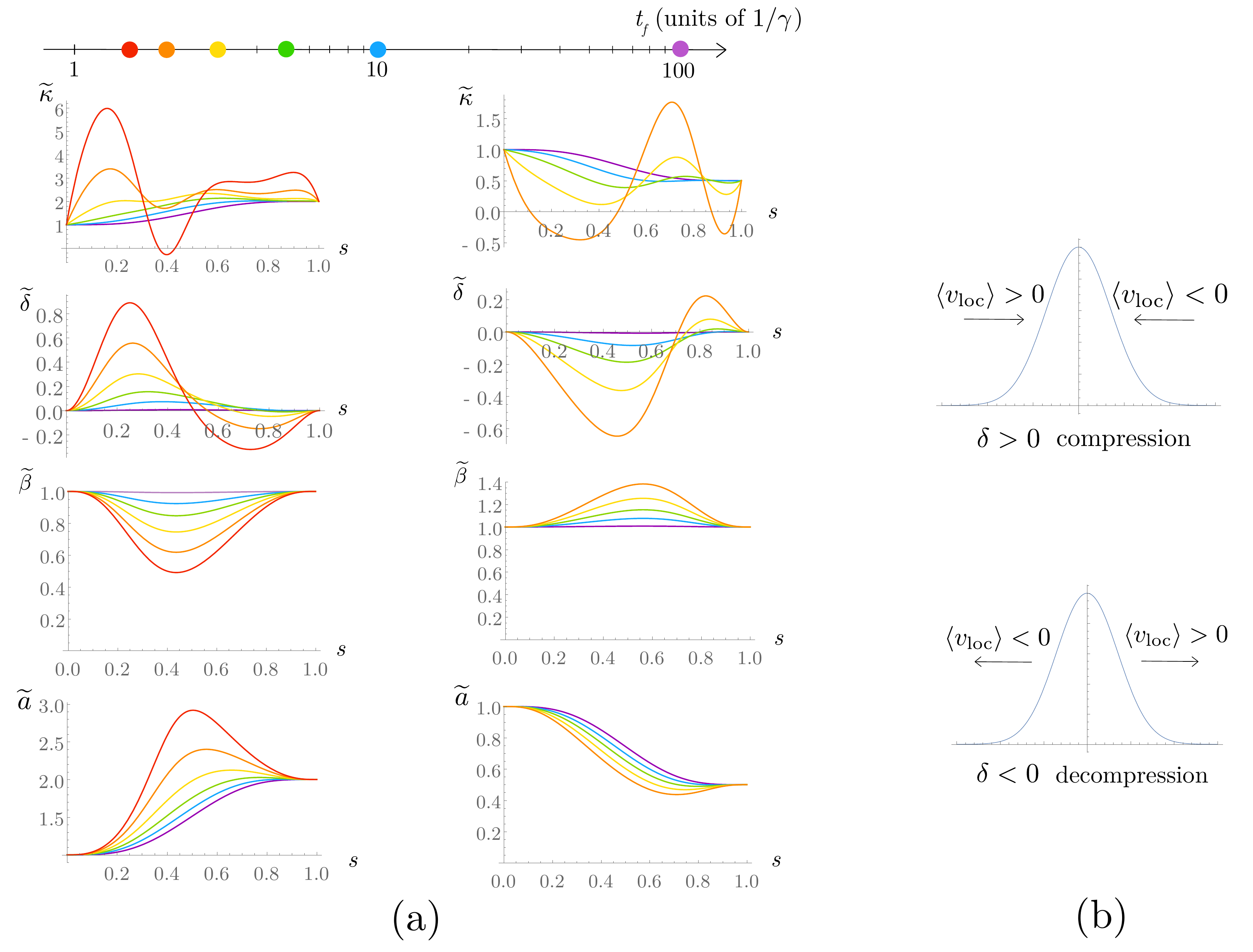}
\caption{(a) Evolution of the functions $\widetilde{\kappa}(s)$, $\widetilde{\delta}(s)$, $\widetilde{\beta}(s)$, $\widetilde{a}(s)$ when the duration $t_f$ of the protocol is decreased, 
thereby moving along the thin black bisectrix in Figs. \ref{phasediag} and \ref{phasediag2} ($\omega_i=\gamma$). The left column is for a compression ($\chi=2$) 
and the right column is for a decompression ($\chi=0.5$). Both columns are for protocol A (with Eq. \eqref{Delta}). The color code is explained in the upper
logarithmic time arrow. The red curves are thus for the faster protocol, that we omitted for decompression, as it does not bring a different information from the orange curve. (b) Illustration of the relation between the sign of $\delta$ and the monotony of $a$. If $\delta$ is positive, the particle density function tends to compress whereas 
when it is negative, it tends to expand.
}
\label{traj}
\end{figure}

We finally address the effect of accelerating the protocol (diminishing $t_f$, other parameters being fixed).
How do the functions \smash{$\widetilde{a}(s)$, $\widetilde{\beta}(s)$, $\widetilde{\delta}(s)$} 
and most importantly the stiffness $\widetilde{\kappa}(s)$ evolve?  The results are reported in Figure~\ref{traj}~(a).
For this ``cut'' across the phase diagrams (line of slope 1), 
we choose $N_{\gamma}=N_{\omega}$, {\it i.e.} $\omega_i=\gamma$ (straight thin black line in Figs. \ref{phasediag} and \ref{phasediag2}). This corresponds to the situation where $t_x=t_v$.

In Fig~\ref{traj}~(a), the purple curves represent a very slow transformation. As expected, the stiffness $\widetilde{\kappa}(s)$ interpolates smoothly between the boundary values, and is well followed by 
the inverse variance of position $\widetilde{a}(s)$: the dynamics is indeed slower than the timescales $t_x$ and $t_v$.
The velocity distribution always remains at equilibrium with the inverse kinetic temperature $\widetilde{\beta}(s)$ that stays close to 1. Velocity and position degrees of freedom 
are decoupled, as shown by a very low crossed term $\widetilde{\delta}(s)$.

When the protocol duration $t_f$ decreases, the inverse kinetic temperature  deviates, temporarily but more and more, from its unit equilibrium value, undergoing a transient heating for a compression 
and a transient cooling for decompression. As explained in paragraph~\ref{sec:exist}, the protocol no longer exists for too fast a compression, 
since $\widetilde{\beta}(s)$ becomes temporarily negative, preventing the ansatz for the particle density function from being well-defined. 
We also observe a change of sign in the crossed term $\widetilde{\delta}$ during the transformation. As this quantity is proportional to the derivative of 
 $\widetilde{a}(s)$ (see equation~\eqref{deltares}), a change of sign transposes into a change of monotony of $\widetilde{a}(s)$, as illustrated in Figure~\ref{traj}~(b). 
This can be understood as follows. We first recast our ansatz~\eqref{ansatz} as
\begin{equation}
K(x,v,t)=N(t) \exp\left(-a x^2-\beta\left( v+ \frac{\delta x}{2 \beta} \right) \right),
\end{equation}
so as to bring out a local mean velocity $\langle v_{\rm loc} \rangle (x)=-\delta x/(2\beta)$. The sign of $\delta$ then indicates the tendency for the local velocity, as sketched in Figure~\ref{traj}~(b). 
This yields a compression if $\delta$ is positive and a decompression if $\delta$ is negative. Indeed, the variance of the position decreases for a positive delta and increases in the opposite case 
(see Figure~\ref{traj}~(a)). 

When $t_f$ diminishes, $\widetilde{a}(s)$ steepens, leading to a pronounced overshoot for a fast compression (resp. undershoot for a fast decompression). 
Finally, the corresponding trap stiffness turns into a complicated non-monotonic function for fast protocols, with negative portions as well as peaked variations, 
as discussed in paragraph~\ref{sec:pbs}. This rather unexpected behavior stems from the fact that controlling the dynamics of both the position and the velocity of the system with 
only one control function, the harmonic force depending on positional degrees of freedom only, is a delicate task. Fig.~\ref{traj}~(a) highlights the difficulty faced when devising an ESE protocol, since even in a case where $t_f$ exceeds the natural relaxation time by a factor three (yellow curves), the driving force and associated response significantly depart from their quasi-static counterpart.

\section{Conclusion}
\label{sec:concl}

In this article, we provide a general framework to study Engineered Swift Equilibration beyond the overdamped regime in which it was initially formulated.
A Brownian particle is here confined in a harmonic potential, the stiffness of which can be changed in time as desired. In addition,
the thermal bath is allowed to have a time dependent temperature $T$. As surprising as this situation might appear,
the latter $T$-control is achievable in the laboratory with, for instance, optically confined colloids \cite{ber14,Chupeau:2018a}.
Trap stiffness and temperature are the two driving 
functions, that need in general to be carefully shaped to meet the desired goal: reaching the target state at the end
of a chosen time $t_f$.
Yet, the formalism becomes cumbersome when $T$ is time dependent, and explicit solutions become elusive. 
ESE techniques being designed especially for experiments and concrete 
applications, it is crucial to be able to exhibit such an explicit protocol. For this reason, we restricted our discussion to harmonic isothermal transport-free transformations, 
that is to say compression and decompression, where the formalism gets significantly simpler. Trap stiffness is thus the only
quantity that is monitored by the experimentalist.

We discussed the explicit and analytical methods that can be employed, and analyzed the corresponding protocols
as a function of the two key quantities of the problem, formed by the ratio between the relevant characteristic timescales. We summarized this analysis 
in two phase diagrams and discussed the range of applicability of our approach, that is ``limitation-free'' in the case of a decompression and limited to protocols 
longer than some lower bound ruled by the friction time in the case of compression. We also investigated core characteristics of our protocol, such as the crossover to the overdamped limit
(where algebra is much simpler). Some attention was also paid to the relevance of the ESE protocol for each set of parameters, leading to investigate the influence of the protocol duration compared to the relaxation timescales on the shape of the trap stiffness. Finally, we discussed the robustness of the phase diagrams presented, by comparing the outcome of
two distinct protocols in a parameter range where rather exotic drivings emerge.

Interesting venues for future work include extending our treatment to baths with time dependent temperature. This additional driving degree of freedom presumably 
leads to more regular protocols. Roughly speaking, the stiffness will ensure the compression or decompression in position 
space while temperature will take care of the velocity degrees of freedom. We thereby expect to overcome the limitations brought to the fore here, 
such as the non-existence of the underlying ansatz (which may become un-normalizable) 
and the odd shape of the stiffness. A first evidence of such an experimental achievement is presented in~\cite{Chupeau:2018a} where an effective modulation of the bath temperature allowed to perform a quick decompression without having to resort to a transiently negative stiffness. Our work also opens interesting perspectives for transformations including transport. Finally, the study of 
non-harmonic driving, energetics, and optimal features appears timely.

\section{Acknowledgments}
This work as been supported by the ERC contract OUTEFLUCOP. 
We acknowledge funding from the
Investissement d'Avenir LabEx PALM program (Grant No.
ANR-10-LABX-0039-PALM). We also thank A. Prados, C. Plata, A. Chepelianskii and O. Dulieu for insightful discussions.

%\bibliographystyle{ieeetr}
%\bibliography{biblio_ESE.bib}

\end{document}